\documentclass[twocolumn]{aastex631}

\usepackage{graphicx} 
\usepackage{footmisc}
\usepackage{makecell}

\sfcode`A=1000 \sfcode`B=1000 \sfcode`C=1000 \sfcode`D=1000
\sfcode`E=1000 \sfcode`F=1000 \sfcode`G=1000 \sfcode`H=1000
\sfcode`I=1000 \sfcode`J=1000 \sfcode`K=1000 \sfcode`L=1000
\sfcode`M=1000 \sfcode`N=1000 \sfcode`O=1000 \sfcode`P=1000
\sfcode`Q=1000 \sfcode`R=1000 \sfcode`S=1000 \sfcode`T=1000
\sfcode`U=1000 \sfcode`V=1000 \sfcode`W=1000 \sfcode`X=1000
\sfcode`Y=1000 \sfcode`Z=1000

\def\hst {{\it HST}}

\def\Bagpipes {\textsc{Bagpipes}}
\def\kms{km~s$^{-1}$}
\def\Oii {\hbox{[\ion{O}{2}]}}
\def\Oiii {\hbox{[\ion{O}{3}]}}
\def\Mgii {\hbox{\ion{Mg}{2}}}
\def\Cii {\hbox{\ion{C}{2}]}}
\newcommand{\Neiii}{\hbox{[\ion{Ne}{3}]}}
\newcommand{\Nev}{\hbox{[\ion{Ne}{5}]}}
\newcommand{\Ciii}{\hbox{\ion{C}{3}]}}
\newcommand{\Sii}{\hbox{\ion{S}{2}}}
\newcommand{\Nii}{\hbox{\ion{N}{2}}}

\def\Hbeta {\hbox{H$\beta$}}

\newcommand{\Msun}{\hbox{M$_\sun$}}

\newcommand{\xc}{{\sc xcsao}}
\newcommand{\no}{\nodata}

\received{January 29, 2025}
\revised{March 6, 2025}
\accepted{March 10, 2025}
\submitjournal{PASP}

\shorttitle{Redshifts near 3C~220.3}
\shortauthors{Hyman, Willner, Wilkes}

\begin{document}

\title{Redshifts of Objects near 3C~220.3}

\author[0000-0002-6036-1858]{S\'oley \'O. Hyman}
\affiliation{Univ.\ of Arizona, Dept.\ of Astronomy, Tucson, AZ 85719, USA}
\affiliation{Center for Astrophysics \textbar\ Harvard \& Smithsonian, Cambridge, MA 02138, USA}
\author[0000-0002-9895-5758]{S.~P.~Willner}
\affiliation{Center for Astrophysics \textbar\ Harvard \& Smithsonian, Cambridge, MA 02138, USA}
\author[0000-0003-1809-2364]{Belinda J. Wilkes}
\affiliation{H.\ H.\ Wills Physics Laboratory, Univ.\ of Bristol, Bristol BS8 1TL, UK}
\affiliation{Center for Astrophysics \textbar\ Harvard \& Smithsonian, Cambridge, MA 02138, USA}

\begin{abstract}
In the course of studying the 3C~220.3 lensing system, spectra were obtained with the Binospec instrument on the MMT for 511 additional objects in 3C~220.3's vicinity. These gave 146 good-quality galaxy redshifts and identified 126 Galactic stars.  The galaxy redshift histogram shows a peak near 3C~220.3's redshift, but there is no evidence for or against a galaxy group within 2~Mpc of 3C~220.3 itself. The spectra revealed 12 AGN candidates including a likely $z\approx4.64$ broad-line QSO. Visible and near-infrared imaging with HST allowed morphological classifications of 14 galaxies. One system is a potential analog of the Milky Way--LMC system with stellar mass ratio $\sim$0.6.
\end{abstract}


\section{Introduction} \label{sec:intro}
3C~220.3 is a narrow-line radio galaxy at redshift $z=0.6850$  that lenses a $z=2.221$ submillimeter galaxy (SMG) into a nearly perfect Einstein ring \citep{Haas_et_al}. \citet{hyman2024} presented new spectra showing that ``galaxy~B,'' near the radio host in projection, is part of the lensing system and likely interacting with the radio host ``galaxy~A.''  The new spectra were obtained using Binospec \citep{Fabricant2019}, a wide-field spectrograph at the MMT. The observations used slit masks that included many objects in addition to the lensing system. This paper reports redshifts and other properties of the additional objects.

One open question is whether the 3C~220.3 radio host is part of a galaxy group or not.  Unfortunately, the current observations do not answer that question, but future observations need not include objects with redshifts reported here.

\section{Observations} \label{sec:observations}
Binospec observations were obtained in 2020 and in 2021 with two different masks having slit position angles of 130\degr\ and 214\degr, respectively. The masks were optimized for the 3C~220.3 system, and additional targets were chosen by position from the NASA/IPAC Extragalactic Database (NED)\footnote{The NASA/IPAC Extragalactic Database (NED) is funded by the National Aeronautics and Space Administration and operated by the California Institute of Technology.} to maximize the number of targets observed.

\begin{figure}[htb]
\includegraphics[width=\columnwidth]{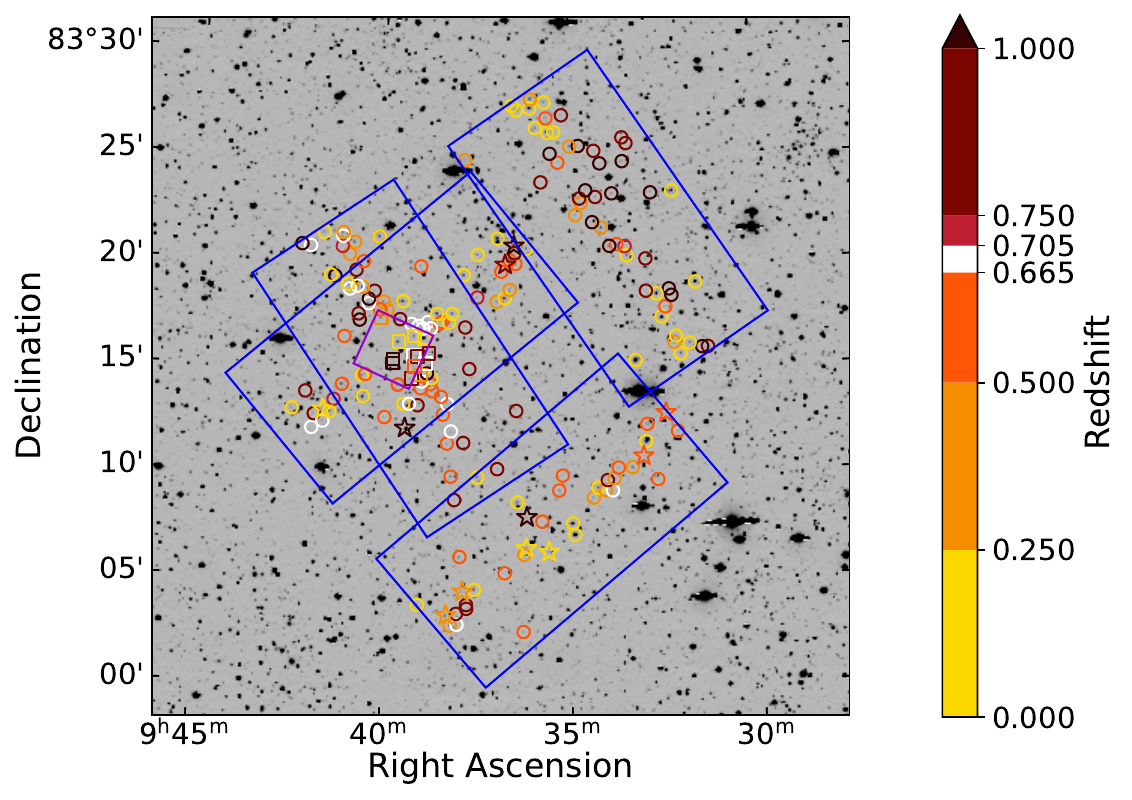}
\caption{Binospec field layout. The background negative image of the field is the coadded $g$+$r$+$z$ MzLS+BASS data from the DECaLS Data Release 10 (credit to Legacy Surveys/D.\ Lang, Perimeter Institute). The magenta square shows the HST footprint, which is centered on 3C~220.3. The blue rectangles indicate the Binospec 8\arcmin$\times$15\arcmin\ fields of view. The spatial dimension of the slit is parallel to the long dimension of each FoV, which corresponds to the position angle (PA) given in the Table~\ref{table:bino_data} notes. Galaxies listed in Tables~\ref{table:binospec-HST-galaxies}--\ref{table:binospec-galaxies} are marked and color-coded by redshift as indicated in the color bar. The white region of the color bar highlights the redshift of the 3C~220.3 system (the primary target of the observations) at $z = 0.685 \pm 0.02$. AGN candidates are indicated by stars, galaxies visible in the \hst\ image are represented by squares, and all other galaxies are indicated by circles.}
\label{fig:gals-sky}
\end{figure}

All Binospec observations used the 270~lpm grating with central wavelength 6560~\AA.  Wavelength coverage was approximately 3900--9240~\AA, depending on each object's position in the field of view, with spectral resolution $R\approx1340$. All slits were 1\farcs0 wide in the dispersion direction and 4\farcs0 to 10\farcs0 long in the spatial direction, except for the 3C~220.3 slit in 2021, which was 20\arcsec\ long.
The 2020 Binospec observations consisted of ten 20-minute exposures with average seeing of 1\farcs08.  The 2021 observations had only four 20-minute exposures, but the average seeing was 0\farcs96. Table~\ref{table:bino_data} lists the  exposure details for both 2020 and 2021, and Figure~\ref{fig:gals-sky} shows the layout of the respective fields of view (FoVs) on the sky. The spectra were processed and reduced with the CfA Binospec pipeline \citep{Kansky2019} v1.0-20190502.

\begin{table}[htb]
\caption{MMT Binospec Observations}
\label{table:bino_data}
\centering
\begin{tabular}{lcchcc}
\hline \hline
Date & UT Start & LST Start & Exposure (s) & Airmass & Seeing (\arcsec) \\
\hline
2020-01-19\tablenotemark{a} & 10:13:30 & 10:43:05 & 1200 & 1.61 & 1.04 \\
 & 10:34:19 & 11:03:57 & 1200 & 1.62 & 1.25 \\
2020-01-24\tablenotemark{a} & 06:10:00 & 06:58:39 & 1200 & 1.67 & 0.89 \\
 & 06:30:51 & 07:19:33 & 1200 & 1.65 & 1.01 \\
 & 06:51:41 & 07:40:25 & 1200 & 1.64 & 1.10 \\
 & 07:12:32 & 08:01:20 & 1200 & 1.63 & 1.19 \\
2020-01-29\tablenotemark{a} & 04:50:38 & 05:58:46 & 1200 & 1.73 & 1.04 \\
 & 05:11:29 & 06:19:39 & 1200 & 1.70 & 1.05 \\
 & 05:32:18 & 06:40:32 & 1200 & 1.68 & 1.17 \\
 & 05:53:05 & 07:01:23 & 1200 & 1.67 & 1.06 \\
2021-02-07\tablenotemark{b} & 10:12:07 & 11:59:35 & 1200 & 1.66 & 0.85 \\
 & 10:32:56 & 12:20:27 & 1200 & 1.67 & 0.99 \\
 & 10:53:44 & 12:41:19 & 1200 & 1.69 & 1.06 \\
 & 11:14:33 & 13:02:12 & 1200 & 1.71 & 0.93 \\
\hline
\end{tabular}
\raggedright
\tablecomments{All exposure times were 1200~s. Seeing was measured by the wavefront-sensor camera. Weather was clear for all exposures.}
\tablenotetext{a}{Program SAO-8-20a, PI Hyman, mask \texttt{Hyman2020A-3C220p3s\_563}, PA 130\degr.}
\tablenotetext{b}{Program SAO-7-21a, PI Hyman, mask \texttt{3C220p3c\_571}, PA $-146$\degr.}
\end{table}

\begin{table}[htb]
\caption{Summary of object classifications}
\label{table:object-counts}
\centering
\begin{tabular}{lccc}
\hline \hline
 & 2020 & 2021 & Duplicate targets\tablenotemark{a}\\
\hline
Galaxies & 84 & 112 & 5 (2)\\
~~~AGN candidates               & 11 &   1 & 0\\
~~~Visible in 2014 \hst\tablenotemark{b} &  9 &  10 & 4 (2)\\
~~~Other galaxies               & 65 & 101 & 1\\
Stars   &  56 &  70 &  1\\
Unclear & 113 &  69 &  3 (2)\\
Total   & 253 & 251 & 11\\
\hline
\end{tabular}
\raggedright
\tablenotetext{a}{Values in parentheses indicate number of slits that appeared as some sort of galaxy in one year and had no trace or unclear spectra in the other year.}
\tablenotetext{b}{Includes slit of 3C~220.3. See \citet{hyman2024} for a detailed analysis of the spectral data. None of the other galaxies in the HST image were classified as AGN.}
\end{table}

Each spectrum was analyzed in three complementary ways: using \xc, using SpecPro \citep{specpro}, and by visual examination of the two-dimensional (2D) spectra in DS9 \citep{ds9}.  \xc\ is an automated and quantitative technique based on cross correlation \citep{Kurtz1998}, but it analyzes only the 1D spectrum produced by the data pipeline.  Visual examination showed which spectra had no signal and sometimes revealed bright emission lines that \xc\ had not found. SpecPro has intermediate characteristics: less quantitative than \xc\ but displays the 1D and 2D spectra and allows a choice of which pixels in the spatial ($y$) direction to fit. SpecPro classifications and redshifts were based on the provided spectral templates (\citeauthor[][their Table~1]{specpro}). When there are multiple traces in a single 2D spectrum, they are labeled with the $y$ coordinate, and the redshift for any $y$ not found by the pipeline came from SpecPro.

\begin{figure*}[h]
\includegraphics[width=\textwidth]{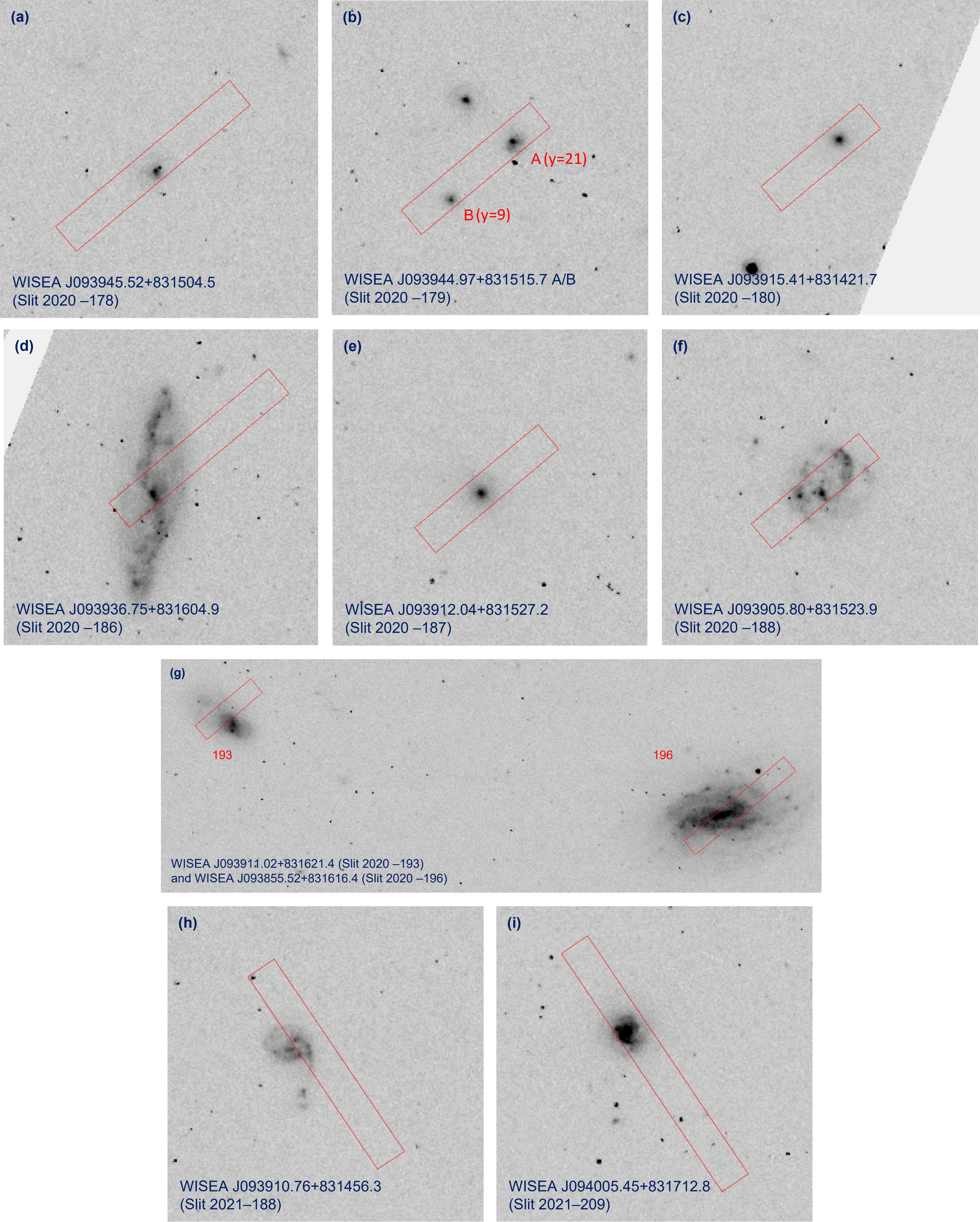}
\caption{2014 \hst \ visible light data cutouts (combined F606W/F814W, log scale) of sources in Table~\ref{table:binospec-HST-galaxies} with Binospec slits overplotted in red. Cutouts are $10\arcsec \times 10\arcsec$ except for (g), which is $37\arcsec \times 13\arcsec$. Slits are 1\arcsec \ wide, with varying lengths (see Section~\ref{sec:observations}). North is up and east is to the left. All sources are within the 2014 \hst\ WFC3-UVIS field of view.}
\label{fig:HST-UVIS-gal-cutouts}
\end{figure*}

\begin{figure*}[ht]
\includegraphics[width=\textwidth]{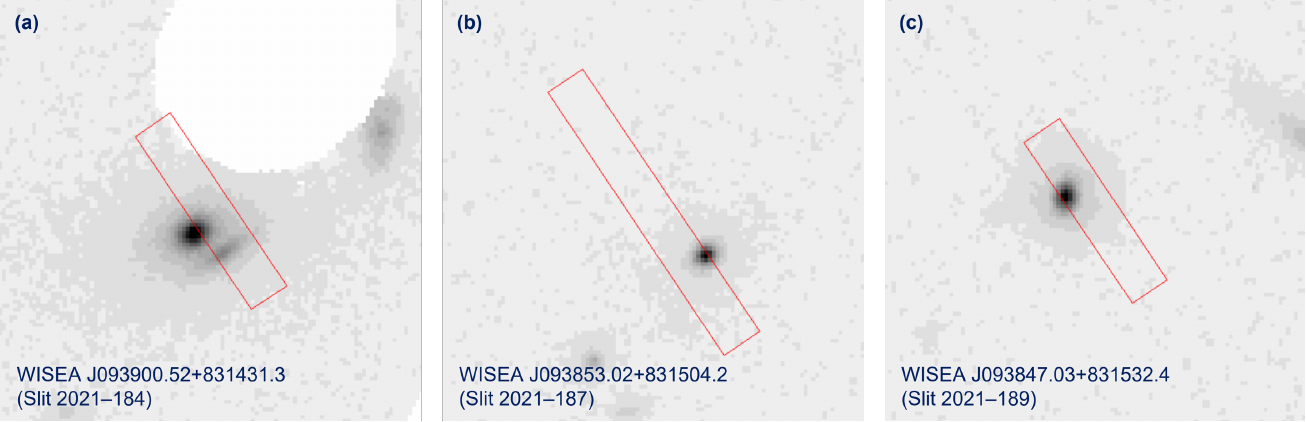}
\caption{2014 \hst \ IR data cutouts (log scale) of sources in Table~\ref{table:binospec-HST-galaxies} with Binospec slits overplotted in red. Cutouts are $10\arcsec \times 10\arcsec$. Slits are 1\arcsec\ wide, with varying lengths (see Section~\ref{sec:observations}). North is up and east is to the left. All sources are within the 2014 \hst\ WFC3-IR field of view.}
\label{fig:HST-IR-gal-cutouts}
\end{figure*}

Objects with sufficient signal to distinguish a trace, continuum, or emission features were classified as galaxies or stars, and approximate spectral types for the stars were assigned according to SpecPro templates.
The galaxy spectra were divided into galaxies visible in the 2014 \hst\ data \citep{hyman2024}, candidate active galactic nuclei (AGN---none of them within the HST coverage), and all other galaxies. Galaxies that fell within the 2014 \hst\ field of view (Table~\ref{table:binospec-HST-galaxies}) were also assigned a galaxy morphology (spiral or elliptical) based on visual evaluation, and Figures~\ref{fig:HST-UVIS-gal-cutouts} and~\ref{fig:HST-IR-gal-cutouts} show the corresponding cutout images and slit placements. AGN candidates were classified
(Table~\ref{table:binospec-agn-candidates}) based on broad or high-excitation emission lines.
Table~\ref{table:object-counts} summarizes these classifications, and Tables~\ref{table:binospec-HST-galaxies}, 
\ref{table:binospec-agn-candidates},
\ref{table:binospec-galaxies}, and~\ref{table:binospec-stars} give redshifts and other information for, respectively, galaxies visible in the 2014 \hst\ data,
AGN candidates, all galaxies and non-detections, and stars.

% \begin{table}[ht]
% \caption{Summary of object classifications}
% \label{table:object-counts}
% \centering
% \begin{tabular}{lccc}
% \hline \hline
%  & 2020 & 2021 & Duplicate targets\tablenotemark{a}\\
% \hline
% Galaxies & 84 & 112 & 5 (2)\\
% ~~~AGN candidates               & 11 &   1 & 0\\
% ~~~Visible in 2014 \hst\tablenotemark{b} &  9 &  10 & 4 (2)\\
% ~~~Other galaxies               & 65 & 101 & 1\\
% Stars   &  56 &  70 &  1\\
% Unclear & 113 &  69 &  3 (2)\\
% Total   & 253 & 251 & 11\\
% \hline
% \end{tabular}
% \raggedright
% \tablenotetext{a}{Values in parentheses indicate number of slits that appeared as some sort of galaxy in one year and had no trace or unclear spectra in the other year.}
% \tablenotetext{b}{Includes slit of 3C~220.3. See \citet{hyman2024} for a detailed analysis of the spectral data. None of the other galaxies in the HST image were classified as AGN.}
% \end{table}

\begin{figure}[htb]
\includegraphics[width=\columnwidth,clip=true,trim=0 2 0 1]{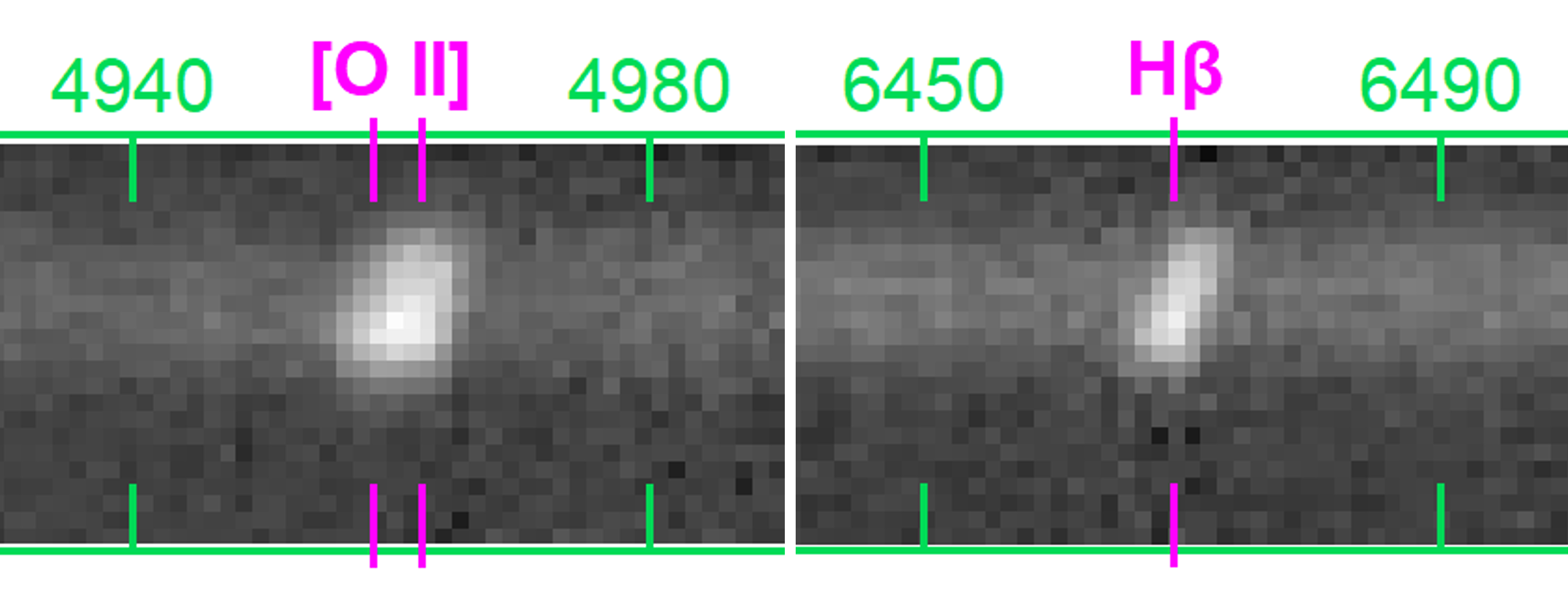}
\caption{2D spectrum closeups of the \Oii\ doublet and \Hbeta\ line in a galaxy at $z=0.3308$ (slit 2021.210).  Observed wavelength increases left to right as shown by the green labels, and the $y$ (spatial) coordinate is vertical. Magenta ticks mark the rest wavelengths of the \Oii\ and H$\beta$ lines. The breadth of the \Oii\ doublet is apparent. A spectrum showing only a single emission feature with this width would have $Q=2$, but this object has $Q=4$ because additional lines confirm the redshift.}
\label{fig:q-ex}
\end{figure}

The tables of galaxy redshifts (Tables~\ref{table:binospec-HST-galaxies}--\ref{table:binospec-galaxies}) give a redshift quality rating $Q$, similar to those used by
other studies \citep[e.g.,][]{Colless2001,Jones2009,Mahony2010,Newman2013,  
Healy2021}. While the rating is subjective, it is based on specific criteria as follows:
\begin{itemize}
\vspace{-1.5ex}
\setlength{\itemsep}{-\parsep}
    \item[4]--- definitive redshift based on unambiguous identifications of multiple features.
    \item[3]--- good redshift based on multiple features but minor doubt about feature identification or reality (e.g., S/N not quite sufficient for $Q=4$).
    \item[2]--- likely redshift but not certain (e.g., not enough features, features could be noise, etc.).  Binospec (barely) resolves the \Oii\ doublet (Figure~\ref{fig:q-ex}), especially at $z\ga0.3$, and a resolved emission line with no other clear spectral features was assumed to be \Oii\ with the resulting redshift assigned $Q=2$.
    \item[1]--- vague redshift indication (e.g., single, unidentified emission line or possible features that might be noise).
    \item[0]--- no redshift, often no signal at all.
\end{itemize}

\section{Results}
   
\subsection{Redshift Distribution}
As expected for the source selection, nearly all the galaxies have $z\le1$.  Only six of 181 $Q\ge2$ galaxies have $z>1$, and one of those is the lensed SMG behind 3C~220.3 \citep[$z=2.221$;][]{Haas_et_al,hyman2024}. Four of the five other $z>1$ sources have QSO spectra.  Figure~\ref{fig:z-dist} shows the redshift distribution of all galaxies. There are numerous peaks along the line of sight indicating large-scale structure.

\subsection{Potential Group Members}
\begin{table}
\caption{Possible 3C 220.3 Companions}
\label{t:group}
\begin{tabular}{cRrr}
\hline\hline
Slit & $\Delta V$\rlap{\tablenotemark{a}} & ~$Q$ & 
$D$/kpc\rlap{\tablenotemark{b}}\\
\hline
2021.168&   62& 3 & 1320\\
2021.177&   -2& 4 & 1000\\
2021.182& -126& 4 & 610\\
2021.225& -553& 3 & 1770\\
%2021.244& -550& 3 & 2890\\
\hline
\end{tabular}
\tablenotetext{a}{Rest-frame ($z=0.685$) velocity difference from 3C~220.3 in \kms.}
\tablenotetext{b}{Projected separation from 3C~220.3 in kpc based on angular scale  7.312\,kpc\,arcsec$^{-1}$ from the same cosmology as adopted by \citet{hyman2024}.}
\end{table}

\begin{figure}[htb]\includegraphics[width=\columnwidth]
{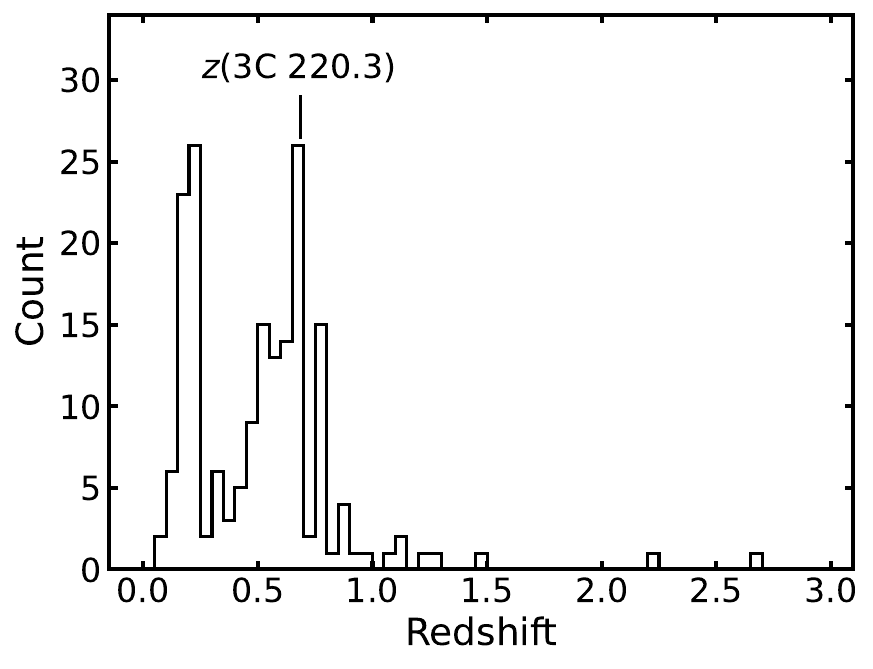}
\caption{Histogram of galaxy redshifts with redshift quality $Q \geq 2$. The 3C~220.3 $z=0.6850$ \citep{hyman2024} is marked.}
\label{fig:z-dist}
\end{figure}

Although the redshift histogram (Figure~\ref{fig:z-dist}) shows a narrow peak at 3C~220.3's redshift, only 8 of the 30 galaxies in the peak
are within (projected) 1 Mpc of 3C~220.3's position. A wide search for potential members of a group associated with 3C~220.3 found only four candidates in the current data. Table~\ref{t:group} lists the galaxies with projected separation $<$2~Mpc, rest-frame velocity difference $<$1000~\kms, and $Q\ge2$. The only strong candidate, in slit 2021.182, has a small velocity difference from 3C~220.3 but a wide separation for a group member.  Given the sparseness of the existing redshifts, the inability to place slits on galaxies near 3C~220.3, and the lack of redshift pre-selection, the search neither confirms nor refutes the presence of a galaxy group. However, based on Figure~\ref{fig:z-dist}, 3C~220.3 does appear to be part of a large-scale over-density near its redshift.

Figure~\ref{fig:gals-sky} shows five closely spaced galaxies near 3C~220.3's redshift just outside the HST footprint. All five are at nearly the same redshift $\langle z\rangle=0.6749$ with a rest-frame velocity dispersion of 236~\kms. This suggests they are members of a galaxy group or small cluster. The five are centered about 1\farcm6 or (projected) 720~kpc from 3C~220.3 with a rest-frame velocity difference of $-1800$~\kms. The separation and velocity difference are too large for 3C~220.3 to be a member of the putative group, but the proximity of the group to 3C~220.3 is further indication of a large-scale overdensity in the region. One of the group members (WISEA J093904.38+831650.8, slit 2020.197) matches the X-ray source 2CXO J093904.3+831650. There is a brighter X-ray source, 2CXO J093911.2+831652, in the midst of this group, but its redshift is unknown.

\subsection{Potential Milky Way--LMC Analog}
Interestingly, the galaxies in slits 2020.193/2021.196 and 2020.196/2021.193 (Table~\ref{table:binospec-HST-galaxies}, Figure~\ref{fig:HST-UVIS-gal-cutouts}g) are at almost the same redshift ($z\approx 0.10$) and have a projected separation of approximately 27\farcs8. This corresponds to a physical distance of 53~kpc, which is similar to the distance between our Milky Way (MW) and the Large Magellanic/Milky Cloud (LMC) of $\sim$50~kpc. Both galaxies are within the \hst\ coverage, with 2020.196/2021.193 appearing to be a large spiral, while 2020.193/2021.196 appears smaller and less defined, with a possible bar or spiral feature. Given these properties, this system could perhaps be a distant analog to our own. 

Spectral energy density (SED) fitting of the two galaxies was done using \Bagpipes\ \citep{2018MNRAS.480.4379C} with the redshifts in Table~\ref{table:binospec-HST-galaxies}, a delayed exponential star formation history model \citep{1986A&A...161...89S,2002ApJ...576..135G_delSFH}, \citet{2002MNRAS.336.1188K} stellar initial mass function (IMF), and \citet{dust_Calzetti_2000} dust attenuation law. The photometric data points came from the 2014 \hst\ F606W, F814W, and F160W imaging and was supplemented by photometry in WISE passbands W1, W2, and W3 from \citet{2013wise.rept....1C} as listed in NED. The resulting fits give an stellar mass estimate of $6.3^{+1.6}_{-1.3} \times 10^{10} ~ \Msun$ for the large spiral and $3.7^{+1.0}_{-0.9} \times 10^{10} ~ \Msun$ for the secondary galaxy, i.e., a mass ratio of $\sim$0.6 compared to $\sim$0.25 for the LMC/MW system \citep{Koposov2023}.

\section{Summary}

MMT/Binospec spectra of sources in the vicinity of the 3C~220.3 lensing system provide good-quality  redshifts for 146 galaxies, including 12 AGN, and identifications of 126 Galactic stars.
The redshift distribution places 3C~220.3 in a large-scale over-density around its redshift.  This over-density also contains a group or small cluster 720 kpc (projected) from 3C~220.3 with velocity difference $-$1800~\kms, but the separation is too large for 3C~220.3 to be a member of this group. The evidence for a group or cluster around 3C~220.3 itself is inconclusive, and spectroscopy of more galaxies near the system is needed. 

Modern multi-object spectrographs observe a plethora of targets with each exposure. It is not always obvious which results will be useful, but results should be made publicly available.

\begin{acknowledgments}
This paper uses data products produced by the OIR Telescope Data Center, supported by the Smithsonian Astrophysical Observatory. The authors especially thank Sean Moran for maintaining the Binospec data pipeline and for combining the 2020 spectra for us.

B.J.W. acknowledges the support of the Royal Society and the Wolfson Foundation at the University of Bristol

This research made use of SAOImage DS9, developed by Smithsonian Astrophysical Observatory \citep{2000ascl.soft03002S,ds9}, IRAF \citep{1986SPIE..627..733T,1993ASPC...52..173T,1999ascl.soft11002N}, SpecPro \citep{specpro}, Photutils, an Astropy package for detection and photometry of astronomical sources \citep{larry_bradley_2023_7946442} Regions, an Astropy package for region handling \citep{larry_bradley_2022_7259631}.

% NED acknowledgement: https://ned.ipac.caltech.edu/Documents/Overview/
This research has made use of the NASA/IPAC Extragalactic Database (NED), which is funded by the National Aeronautics and Space Administration and operated by the California Institute of Technology.

This publication makes use of data products from the Wide-field Infrared Survey Explorer, which is a joint project of the University of California, Los Angeles, and the Jet Propulsion Laboratory/California Institute of Technology, funded by the National Aeronautics and Space Administration.

We respectfully acknowledge the University of Arizona is on the land and territories of Indigenous peoples. Today, Arizona is home to 22 federally recognized tribes, with Tucson being home to the O’odham and the Yaqui. Committed to diversity and inclusion, the University strives to build sustainable relationships with sovereign Native Nations and Indigenous communities through education offerings, partnerships, and community service.

The \hst\ data used in this paper can be accessed at \dataset[10.17909/ghjp-t523]{\doi{10.17909/ghjp-t523}}.
\end{acknowledgments}

%%%%%%%%%%%%%%%%%%%%%%%%%%%%%%%%%%%%%%%%%%%%%%%
% Software
\software{SAOImageDS9 \citep{2000ascl.soft03002S,ds9}, SpecPro \citep{specpro}, Binospec Software System \citep{Kansky2019}, IRAF \citep{1986SPIE..627..733T,1993ASPC...52..173T,1999ascl.soft11002N}, Astropy \citep{astropy:2013,astropy:2018,astropy:2022}, Bagpipes \citep{2018MNRAS.480.4379C}, Photutils \citep{larry_bradley_2023_7946442}, Regions \citep{larry_bradley_2022_7259631}, Pyregion, NumPy \citep{harris2020array}, Matplotlib \citep{Hunter:2007}}
%%%%%%%%%%%%%%%%%%%%%%%%%%%%%%%%%%%%%%%%%%%%%%%

\startlongtable
\begin{deluxetable*}{lllcll}
\tabletypesize{\footnotesize}
\tablecaption{Galaxies visible in 2014 \hst\ data and classified with Binospec observations}
\label{table:binospec-HST-galaxies}
\tablehead{\colhead{Slit} & \colhead{WISEA Object Name} & \colhead{$z$} & \colhead{$Q$} & \colhead{Spectrum template} & \colhead{\hst\ morphology}}
\startdata
2020.178 & J093945.52+831504.5 & 1.081 & 1 & unclear & faint face-on spiral \\
2020.179A & J093944.97+831515.7A & 0.7892 & 3 & spiral & possible face-on spiral \\
2020.179B & J093944.97+831515.7B & \nodata & 0 & unclear & face-on spiral \\
2020.180 & J093915.41+831421.7 & 0.791 & 1 & likely elliptical & diffuse source \\
2020.186 & J093936.75+831604.9 & 0.2441 & 4 & spiral & large inclined spiral with warped disk \\
2021.198 & same & \\
2020.187 & J093912.04+831527.2 & 0.680 & 1 & likely elliptical & diffuse source \\
2020.188 & J093905.80+831523.9 & 0.7915 & 1 & spiral & face-on spiral \\
2021.191 & same & \\
2020.193 & J093911.02+831621.4 & 0.1006 & 4 & spiral & elliptical or inclined spiral (no visible arms) \\
2021.196 & same & \\
2020.196 & J093855.52+831616.4 & 0.101 & 4 & spiral & inclined spiral galaxy \\
2021.193 & same &\\
2021.184 & J093900.52+831431.3 & 0.3245 & 2 & spiral & seen in HST IR \\
2021.187 & J093853.02+831504.2 & 0.67 & 3 & spiral & seen in HST IR \\
2021.188 & J093910.76+831456.3 & 0.578 & 3 & spiral & face-on spiral \\
2021.189 & J093847.03+831532.4 & 0.793 & 2 & early spiral & seen in HST IR \\
2021.209 & J094005.45+831712.8 & 0.399 & 3 & spiral & possible face-on spiral \\
\enddata
\end{deluxetable*}

\startlongtable
\begin{deluxetable*}{llccl}
\tabletypesize{\footnotesize}
\tablecaption{AGN candidates classified with Binospec observations}
\label{table:binospec-agn-candidates}
\tablehead{\colhead{Slit} & \colhead{Object Name} & \colhead{$z$} & \colhead{$Q$} & \colhead{Potential classification}}
\startdata
2020.007 & WISEA J093818.00+830309.9 & 0.4438 & 4 & NL AGN or star-forming (SF) galaxy \\
2020.017 & GALEXASC J093752.27+830416.9 & 0.4997 & 4 & SF (low metallicity) or NL AGN \\
2020.048 & GALEXASC J093611.04+830620.6 & 0.1557 & 4 & NL AGN or metal-poor SF galaxy; extended/irregular in SDSS images \\
2020.054 & GALEXASC J093535.37+830610.0 & 0.1585 & 4 & NL AGN \\
2020.056 & WISEA J093610.95+830748.2 & 1.455 & 4 & BL AGN or quasar; pointlike in SDSS images \\
2020.107 & WISEA J093303.64+831040.1 & 0.5915 & 4 & SF or NL AGN \\
2020.125 & GALEXASC J093226.94+831242.1 & 0.5530 & 4 & NL AGN or SF galaxy \\
2020.144 & WISEA J094136.91+831251.6 & 0.2079 & 4 & NL AGN or SF galaxy \\
2020.149 &WISEA J094119.78+831320.6& 4.64 & 1 & See table note \textit{a}  \\
2020.241 & WISEA J093644.51+831945.4 & 0.9455 & 4 & BL AGN or quasar \\
2020.249 & WISEA J093630.46+832038.8 & 1.2647 & 4 & BL AGN or quasar; pointlike in SDSS images \\
2021.170 & CXOGSG J093925.2+831201 & 2.7 & 3 & BL AGN or quasar \\
\enddata
\tablenotetext{a}{The  spectrum shows a strong break near 6840\,\AA. This is in the atmospheric B band, making the exact wavelength uncertain.  If this is the Balmer break (at $z\approx0.72$), the Ca K and H absorption bands and maybe \Oii\ emission should be visible redward of the break, but none of these are seen. The more likely possibility is a Lyman break at $z\approx4.64$.}
\end{deluxetable*}%


\begin{thebibliography}{}
\expandafter\ifx\csname natexlab\endcsname\relax\def\natexlab#1{#1}\fi
\providecommand{\url}[1]{\href{#1}{#1}}
\providecommand{\dodoi}[1]{doi:~\href{http://doi.org/#1}{\nolinkurl{#1}}}
\providecommand{\doeprint}[1]{\href{http://ascl.net/#1}{\nolinkurl{http://ascl.net/#1}}}
\providecommand{\doarXiv}[1]{\href{https://arxiv.org/abs/#1}{\nolinkurl{https://arxiv.org/abs/#1}}}

\bibitem[{{Astropy Collaboration} {et~al.}(2013){Astropy Collaboration}, {Robitaille}, {Tollerud}, {Greenfield}, {Droettboom}, {Bray}, {Aldcroft}, {Davis}, {Ginsburg}, {Price-Whelan}, {Kerzendorf}, {Conley}, {Crighton}, {Barbary}, {Muna}, {Ferguson}, {Grollier}, {Parikh}, {Nair}, {Unther}, {Deil}, {Woillez}, {Conseil}, {Kramer}, {Turner}, {Singer}, {Fox}, {Weaver}, {Zabalza}, {Edwards}, {Azalee Bostroem}, {Burke}, {Casey}, {Crawford}, {Dencheva}, {Ely}, {Jenness}, {Labrie}, {Lim}, {Pierfederici}, {Pontzen}, {Ptak}, {Refsdal}, {Servillat}, \& {Streicher}}]{astropy:2013}
{Astropy Collaboration}, {Robitaille}, T.~P., {Tollerud}, E.~J., {et~al.} 2013, \aap, 558, A33, \dodoi{10.1051/0004-6361/201322068}

\bibitem[{{Astropy Collaboration} {et~al.}(2018){Astropy Collaboration}, {Price-Whelan}, {Sip{\H{o}}cz}, {G{\"u}nther}, {Lim}, {Crawford}, {Conseil}, {Shupe}, {Craig}, {Dencheva}, {Ginsburg}, {VanderPlas}, {Bradley}, {P{\'e}rez-Su{\'a}rez}, {de Val-Borro}, {Aldcroft}, {Cruz}, {Robitaille}, {Tollerud}, {Ardelean}, {Babej}, {Bach}, {Bachetti}, {Bakanov}, {Bamford}, {Barentsen}, {Barmby}, {Baumbach}, {Berry}, {Biscani}, {Boquien}, {Bostroem}, {Bouma}, {Brammer}, {Bray}, {Breytenbach}, {Buddelmeijer}, {Burke}, {Calderone}, {Cano Rodr{\'\i}guez}, {Cara}, {Cardoso}, {Cheedella}, {Copin}, {Corrales}, {Crichton}, {D'Avella}, {Deil}, {Depagne}, {Dietrich}, {Donath}, {Droettboom}, {Earl}, {Erben}, {Fabbro}, {Ferreira}, {Finethy}, {Fox}, {Garrison}, {Gibbons}, {Goldstein}, {Gommers}, {Greco}, {Greenfield}, {Groener}, {Grollier}, {Hagen}, {Hirst}, {Homeier}, {Horton}, {Hosseinzadeh}, {Hu}, {Hunkeler}, {Ivezi{\'c}}, {Jain}, {Jenness}, {Kanarek}, {Kendrew}, {Kern}, {Kerzendorf}, {Khvalko}, {King}, {Kirkby}, {Kulkarni},
  {Kumar}, {Lee}, {Lenz}, {Littlefair}, {Ma}, {Macleod}, {Mastropietro}, {McCully}, {Montagnac}, {Morris}, {Mueller}, {Mumford}, {Muna}, {Murphy}, {Nelson}, {Nguyen}, {Ninan}, {N{\"o}the}, {Ogaz}, {Oh}, {Parejko}, {Parley}, {Pascual}, {Patil}, {Patil}, {Plunkett}, {Prochaska}, {Rastogi}, {Reddy Janga}, {Sabater}, {Sakurikar}, {Seifert}, {Sherbert}, {Sherwood-Taylor}, {Shih}, {Sick}, {Silbiger}, {Singanamalla}, {Singer}, {Sladen}, {Sooley}, {Sornarajah}, {Streicher}, {Teuben}, {Thomas}, {Tremblay}, {Turner}, {Terr{\'o}n}, {van Kerkwijk}, {de la Vega}, {Watkins}, {Weaver}, {Whitmore}, {Woillez}, {Zabalza}, \& {Astropy Contributors}}]{astropy:2018}
{Astropy Collaboration}, {Price-Whelan}, A.~M., {Sip{\H{o}}cz}, B.~M., {et~al.} 2018, \aj, 156, 123, \dodoi{10.3847/1538-3881/aabc4f}

\bibitem[{{Astropy Collaboration} {et~al.}(2022){Astropy Collaboration}, {Price-Whelan}, {Lim}, {Earl}, {Starkman}, {Bradley}, {Shupe}, {Patil}, {Corrales}, {Brasseur}, {N{\"o}the}, {Donath}, {Tollerud}, {Morris}, {Ginsburg}, {Vaher}, {Weaver}, {Tocknell}, {Jamieson}, {van Kerkwijk}, {Robitaille}, {Merry}, {Bachetti}, {G{\"u}nther}, {Aldcroft}, {Alvarado-Montes}, {Archibald}, {B{\'o}di}, {Bapat}, {Barentsen}, {Baz{\'a}n}, {Biswas}, {Boquien}, {Burke}, {Cara}, {Cara}, {Conroy}, {Conseil}, {Craig}, {Cross}, {Cruz}, {D'Eugenio}, {Dencheva}, {Devillepoix}, {Dietrich}, {Eigenbrot}, {Erben}, {Ferreira}, {Foreman-Mackey}, {Fox}, {Freij}, {Garg}, {Geda}, {Glattly}, {Gondhalekar}, {Gordon}, {Grant}, {Greenfield}, {Groener}, {Guest}, {Gurovich}, {Handberg}, {Hart}, {Hatfield-Dodds}, {Homeier}, {Hosseinzadeh}, {Jenness}, {Jones}, {Joseph}, {Kalmbach}, {Karamehmetoglu}, {Ka{\l}uszy{\'n}ski}, {Kelley}, {Kern}, {Kerzendorf}, {Koch}, {Kulumani}, {Lee}, {Ly}, {Ma}, {MacBride}, {Maljaars}, {Muna}, {Murphy}, {Norman},
  {O'Steen}, {Oman}, {Pacifici}, {Pascual}, {Pascual-Granado}, {Patil}, {Perren}, {Pickering}, {Rastogi}, {Roulston}, {Ryan}, {Rykoff}, {Sabater}, {Sakurikar}, {Salgado}, {Sanghi}, {Saunders}, {Savchenko}, {Schwardt}, {Seifert-Eckert}, {Shih}, {Jain}, {Shukla}, {Sick}, {Simpson}, {Singanamalla}, {Singer}, {Singhal}, {Sinha}, {Sip{\H{o}}cz}, {Spitler}, {Stansby}, {Streicher}, {{\v{S}}umak}, {Swinbank}, {Taranu}, {Tewary}, {Tremblay}, {de Val-Borro}, {Van Kooten}, {Vasovi{\'c}}, {Verma}, {de Miranda Cardoso}, {Williams}, {Wilson}, {Winkel}, {Wood-Vasey}, {Xue}, {Yoachim}, {Zhang}, {Zonca}, \& {Astropy Project Contributors}}]{astropy:2022}
{Astropy Collaboration}, {Price-Whelan}, A.~M., {Lim}, P.~L., {et~al.} 2022, \apj, 935, 167, \dodoi{10.3847/1538-4357/ac7c74}

\bibitem[{Bradley {et~al.}(2022)Bradley, Deil, Ginsburg, Patra, Robitaille, Sipőcz, King, Lim, Homeier, Singer, de~Val-Borro, Jenness, Baumann, Gondhalekar, Donath, Tollerud, Lee, Leinweber, \& Vinícius}]{larry_bradley_2022_7259631}
Bradley, L., Deil, C., Ginsburg, A., {et~al.} 2022, astropy/regions: v0.7, v0.7,  Zenodo, \dodoi{10.5281/zenodo.7259631}

\bibitem[{Bradley {et~al.}(2023)Bradley, Sip{\H o}cz, Robitaille, Tollerud, Vin{\'{\i}}cius, Deil, Barbary, Wilson, Busko, Donath, G{\"u}nther, Cara, Lim, Me{\ss}linger, Conseil, Bostroem, Droettboom, Bray, Bratholm, Barentsen, Craig, Rathi, Pascual, Perren, Georgiev, de~Val-Borro, Kerzendorf, Bach, Quint, \& Souchereau}]{larry_bradley_2023_7946442}
Bradley, L., Sip{\H o}cz, B., Robitaille, T., {et~al.} 2023, astropy/photutils: 1.8.0, 1.8.0,  Zenodo, \dodoi{10.5281/zenodo.7946442}

\bibitem[{{Calzetti} {et~al.}(2000){Calzetti}, {Armus}, {Bohlin}, {Kinney}, {Koornneef}, \& {Storchi-Bergmann}}]{dust_Calzetti_2000}
{Calzetti}, D., {Armus}, L., {Bohlin}, R.~C., {et~al.} 2000, \apj, 533, 682, \dodoi{10.1086/308692}

\bibitem[{{Carnall} {et~al.}(2018){Carnall}, {McLure}, {Dunlop}, \& {Dav{\'e}}}]{2018MNRAS.480.4379C}
{Carnall}, A.~C., {McLure}, R.~J., {Dunlop}, J.~S., \& {Dav{\'e}}, R. 2018, \mnras, 480, 4379, \dodoi{10.1093/mnras/sty2169}

\bibitem[{{Colless} {et~al.}(2001){Colless}, {Dalton}, {Maddox}, {Sutherland}, {Norberg}, {Cole}, {Bland-Hawthorn}, {Bridges}, {Cannon}, {Collins}, {Couch}, {Cross}, {Deeley}, {De Propris}, {Driver}, {Efstathiou}, {Ellis}, {Frenk}, {Glazebrook}, {Jackson}, {Lahav}, {Lewis}, {Lumsden}, {Madgwick}, {Peacock}, {Peterson}, {Price}, {Seaborne}, \& {Taylor}}]{Colless2001}
{Colless}, M., {Dalton}, G., {Maddox}, S., {et~al.} 2001, \mnras, 328, 1039, \dodoi{10.1046/j.1365-8711.2001.04902.x}

\bibitem[{{Cutri} {et~al.}(2013){Cutri}, {Wright}, {Conrow}, {Fowler}, {Eisenhardt}, {Grillmair}, {Kirkpatrick}, {Masci}, {McCallon}, {Wheelock}, {Fajardo-Acosta}, {Yan}, {Benford}, {Harbut}, {Jarrett}, {Lake}, {Leisawitz}, {Ressler}, {Stanford}, {Tsai}, {Liu}, {Helou}, {Mainzer}, {Gettings}, {Gonzalez}, {Hoffman}, {Marsh}, {Padgett}, {Skrutskie}, {Beck}, {Papin}, \& {Wittman}}]{2013wise.rept....1C}
{Cutri}, R.~M., {Wright}, E.~L., {Conrow}, T., {et~al.} 2013, {Explanatory Supplement to the AllWISE Data Release Products}, Explanatory Supplement to the AllWISE Data Release Products, by R. M. Cutri et al.

\bibitem[{{Fabricant} {et~al.}(2019){Fabricant}, {Fata}, {Epps}, {Gauron}, {Mueller}, {Zajac}, {Amato}, {Barberis}, {Bergner}, {Brennan}, {Brown}, {Chilingarian}, {Geary}, {Kradinov}, {McLeod}, {Smith}, \& {Woods}}]{Fabricant2019}
{Fabricant}, D., {Fata}, R., {Epps}, H., {et~al.} 2019, \pasp, 131, 075004, \dodoi{10.1088/1538-3873/ab1d78}

\bibitem[{{Gavazzi} {et~al.}(2002){Gavazzi}, {Bonfanti}, {Sanvito}, {Boselli}, \& {Scodeggio}}]{2002ApJ...576..135G_delSFH}
{Gavazzi}, G., {Bonfanti}, C., {Sanvito}, G., {Boselli}, A., \& {Scodeggio}, M. 2002, \apj, 576, 135, \dodoi{10.1086/341730}

\bibitem[{{Haas} {et~al.}(2014){Haas}, {Leipski}, {Barthel}, {Wilkes}, {Vegetti}, {Bussmann}, {Willner}, {Westhues}, {Ashby}, {Chini}, {Clements}, {Fassnacht}, {Horesh}, {Klaas}, {Koopmans}, {Kuraszkiewicz}, {Lagattuta}, {Meisenheimer}, {Stern}, \& {Wylezalek}}]{Haas_et_al}
{Haas}, M., {Leipski}, C., {Barthel}, P., {et~al.} 2014, \apj, 790, 46, \dodoi{10.1088/0004-637X/790/1/46}

\bibitem[{Harris {et~al.}(2020)Harris, Millman, van~der Walt, Gommers, Virtanen, Cournapeau, Wieser, Taylor, Berg, Smith, Kern, Picus, Hoyer, van Kerkwijk, Brett, Haldane, del R{\'{i}}o, Wiebe, Peterson, G{\'{e}}rard-Marchant, Sheppard, Reddy, Weckesser, Abbasi, Gohlke, \& Oliphant}]{harris2020array}
Harris, C.~R., Millman, K.~J., van~der Walt, S.~J., {et~al.} 2020, Nature, 585, 357, \dodoi{10.1038/s41586-020-2649-2}

\bibitem[{{Healy} {et~al.}(2021){Healy}, {Willner}, {Verheijen}, \& {Blyth}}]{Healy2021}
{Healy}, J., {Willner}, S.~P., {Verheijen}, M.~A.~W., \& {Blyth}, S.~L. 2021, \aj, 162, 193, \dodoi{10.3847/1538-3881/ac0bc6}

\bibitem[{Hunter(2007)}]{Hunter:2007}
Hunter, J.~D. 2007, Computing in Science \& Engineering, 9, 90, \dodoi{10.1109/MCSE.2007.55}

\bibitem[{{Hyman} {et~al.}(2024){Hyman}, {Wilkes}, {Willner}, {Kuraszkiewicz}, {Azadi}, {Worrall}, {Foord}, {Vegetti}, {Ashby}, {Birkinshaw}, {Fassnacht}, {Haas}, \& {Stern}}]{hyman2024}
{Hyman}, S.~{\'O}., {Wilkes}, B.~J., {Willner}, S.~P., {et~al.} 2024, \apj, 974, 171, \dodoi{10.3847/1538-4357/ad68f7}

\bibitem[{{Jones} {et~al.}(2009){Jones}, {Read}, {Saunders}, {Colless}, {Jarrett}, {Parker}, {Fairall}, {Mauch}, {Sadler}, {Watson}, {Burton}, {Campbell}, {Cass}, {Croom}, {Dawe}, {Fiegert}, {Frankcombe}, {Hartley}, {Huchra}, {James}, {Kirby}, {Lahav}, {Lucey}, {Mamon}, {Moore}, {Peterson}, {Prior}, {Proust}, {Russell}, {Safouris}, {Wakamatsu}, {Westra}, \& {Williams}}]{Jones2009}
{Jones}, D.~H., {Read}, M.~A., {Saunders}, W., {et~al.} 2009, \mnras, 399, 683, \dodoi{10.1111/j.1365-2966.2009.15338.x}

\bibitem[{{Joye} \& {Mandel}(2003)}]{ds9}
{Joye}, W.~A., \& {Mandel}, E. 2003, in Astronomical Society of the Pacific Conference Series, Vol. 295, Astronomical Data Analysis Software and Systems XII, ed. H.~E. {Payne}, R.~I. {Jedrzejewski}, \& R.~N. {Hook}, 489

\bibitem[{{Kansky} {et~al.}(2019){Kansky}, {Chilingarian}, {Fabricant}, {Matthews}, {Moran}, {Paegert}, {Duane Gibson}, {Porter}, \& {Roll}}]{Kansky2019}
{Kansky}, J., {Chilingarian}, I., {Fabricant}, D., {et~al.} 2019, \pasp, 131, 075005, \dodoi{10.1088/1538-3873/ab1ceb}

\bibitem[{{Koposov} {et~al.}(2023){Koposov}, {Erkal}, {Li}, {Da Costa}, {Cullinane}, {Ji}, {Kuehn}, {Lewis}, {Pace}, {Shipp}, {Zucker}, {Bland-Hawthorn}, {Lilleengen}, {Martell}, \& {S5 Collaboration}}]{Koposov2023}
{Koposov}, S.~E., {Erkal}, D., {Li}, T.~S., {et~al.} 2023, \mnras, 521, 4936, \dodoi{10.1093/mnras/stad551}

\bibitem[{{Kroupa} \& {Boily}(2002)}]{2002MNRAS.336.1188K}
{Kroupa}, P., \& {Boily}, C.~M. 2002, \mnras, 336, 1188, \dodoi{10.1046/j.1365-8711.2002.05848.x}

\bibitem[{{Kurtz} \& {Mink}(1998)}]{Kurtz1998}
{Kurtz}, M.~J., \& {Mink}, D.~J. 1998, \pasp, 110, 934, \dodoi{10.1086/316207}

\bibitem[{{Mahony} {et~al.}(2010){Mahony}, {Croom}, {Boyle}, {Edge}, {Mauch}, \& {Sadler}}]{Mahony2010}
{Mahony}, E.~K., {Croom}, S.~M., {Boyle}, B.~J., {et~al.} 2010, \mnras, 401, 1151, \dodoi{10.1111/j.1365-2966.2009.15705.x}

\bibitem[{{Masters} \& {Capak}(2011)}]{specpro}
{Masters}, D., \& {Capak}, P. 2011, \pasp, 123, 638, \dodoi{10.1086/660023}

\bibitem[{{National Optical Astronomy Observatories}(1999)}]{1999ascl.soft11002N}
{National Optical Astronomy Observatories}. 1999, {IRAF: Image Reduction and Analysis Facility}, Astrophysics Source Code Library, record ascl:9911.002

\bibitem[{{Newman} {et~al.}(2013){Newman}, {Cooper}, {Davis}, {Faber}, {Coil}, {Guhathakurta}, {Koo}, {Phillips}, {Conroy}, {Dutton}, {Finkbeiner}, {Gerke}, {Rosario}, {Weiner}, {Willmer}, {Yan}, {Harker}, {Kassin}, {Konidaris}, {Lai}, {Madgwick}, {Noeske}, {Wirth}, {Connolly}, {Kaiser}, {Kirby}, {Lemaux}, {Lin}, {Lotz}, {Luppino}, {Marinoni}, {Matthews}, {Metevier}, \& {Schiavon}}]{Newman2013}
{Newman}, J.~A., {Cooper}, M.~C., {Davis}, M., {et~al.} 2013, \apjs, 208, 5, \dodoi{10.1088/0067-0049/208/1/5}

\bibitem[{{Sandage}(1986)}]{1986A&A...161...89S}
{Sandage}, A. 1986, \aap, 161, 89

\bibitem[{{Smithsonian Astrophysical Observatory}(2000)}]{2000ascl.soft03002S}
{Smithsonian Astrophysical Observatory}. 2000, {SAOImage DS9: A utility for displaying astronomical images in the X11 window environment}, Astrophysics Source Code Library, record ascl:0003.002

\bibitem[{{Tody}(1986)}]{1986SPIE..627..733T}
{Tody}, D. 1986, in Society of Photo-Optical Instrumentation Engineers (SPIE) Conference Series, Vol. 627, Instrumentation in astronomy VI, ed. D.~L. {Crawford}, 733, \dodoi{10.1117/12.968154}

\bibitem[{{Tody}(1993)}]{1993ASPC...52..173T}
{Tody}, D. 1993, in Astronomical Society of the Pacific Conference Series, Vol.~52, Astronomical Data Analysis Software and Systems II, ed. R.~J. {Hanisch}, R.~J.~V. {Brissenden}, \& J.~{Barnes}, 173

\end{thebibliography}
\end{document}